\documentclass[prl,twocolumn,showpacs,preprintnumbers,amsmath,amssymb]{revtex4}
\usepackage{graphicx}
\usepackage{dcolumn}
\usepackage{bm}
\usepackage{psfig}
\newcommand \be{\begin{eqnarray}}
\newcommand \ee{\end{eqnarray}}

\begin{document}
\title{Femtosecond formation of collective modes 
due to meanfield fluctuations}
\author{K. Morawetz$^{1,2}$ and P. Lipavsk\'y$^{3}$ and M. Schreiber$^{1}$
}
\affiliation{$^1$Institute of Physics, Chemnitz University of Technology, 
09107 Chemnitz, Germany}
\affiliation{
$^2$Max-Planck-Institute for the Physics of Complex
Systems, N{\"o}thnitzer Str. 38, 01187 Dresden, Germany}
\affiliation{$^3$Faculty of Mathematics and Physics, 
Charles University, Ke Karlovu 5, 12116 Prague 2, Czech Republic}
\begin{abstract}
Starting from a quantum kinetic equation including the 
mean field and a conserving relaxation-time approximation we 
derive an analytic formula which describes the time dependence 
of the dielectric function in a plasma created by a short intense 
laser pulse. This formula reproduces universal features of the
formation of collective modes seen in recent experimental 
data of femtosecond spectroscopy. The presented formula offers
 a tremendous simplification for the description of
the formation of quasiparticle features in interacting systems.
Numerical demanding treatments can now be focused on effects
beyond these gross features found here to be describable analytically. 
\end{abstract}
\maketitle
The last ten years have been characterized by an enormous 
activity about ultrafast excitations in semiconductors, clusters, or 
plasmas by ultrashort laser pulses. The femtosecond spectroscopy 
has opened the exciting possibility to observe directly the 
formation of collective modes and quasiparticles in interacting 
many-body systems. This formation is reflected in the time 
dependence of the dielectric function 
\cite{HTBBAL01,HTBL02,HKTBLVHKA05} or terahertz 
emission \cite{HCFJJ04}. For an overview over theoretical 
and experimental work see \cite{AK04,Mo04}. Such ultrafast 
excitations in semiconductors have been satisfactorily 
described  by calculating nonequilibrium Green's functions 
\cite{BVMH98,GBH99}. 
This approach allows one to describe the formation of 
collective modes \cite{VH00,HKTBLVHKA05} and even exciton 
population inversions \cite{KK04}.

The experimental data in semiconductors like GaAs 
\cite{HTBL02} or InP 
\cite{HKTBLVHKA05} reveal similar features. These 
features are explained by several numerically demanding calculations. Since 
some common features are robust, i.e., independent of the 
actual used material parameters, it should be possible to describe them 
by a simple theory. 
The origin of such robust features likely rests in the short- 
or transient-time behavior itself. 
At short times higher-order correlations have no time yet to 
develop, therefore the dynamics is controlled exclusively 
by mean-field forces. 

Based on the mean-field character of the short-time evolution 
we intend to derive a time-dependent response function from 
the mean-field linear response. We will start from a conserving 
relaxation-time approximation which dates back to an idea of 
Mermin \cite{Mer70,D75}. Our aim is a simple analytic formula 
suitable for fits of experimental data.

The electron motion is controlled by the kinetic energy
$\hat {\cal E}$, the 
external perturbation $\hat V^{\rm ext}$ and the induced 
mean-field potential $\hat {V}^{\rm ind}$. The corresponding 
kinetic equation for the one-particle reduced density matrix $\hat \rho$ reads 
\be
\dot {\hat \rho}+i[\hat {\cal E}+\hat V^{\rm ind}+\hat V^{\rm ext},
\hat \rho] ={\hat \rho^{\rm l.e.}-\hat \rho \over \tau}
\label{1}
\ee
with the relaxation time $\tau$ and a local equilibrium density matrix $\hat \rho^{\rm l.e.}$ determined in the following.
We will calculate
the response in the momentum representation
\be
f({p,q},t)=\left\langle p+{q\over 2}\right|\hat \rho
\left|p-{q\over 2}\right\rangle ,
\label{1a}
\ee
where $\left|p\right\rangle$ is an eigenstate
of momentum $p$. For interpretations it is
more convenient to transform the reduced density 
matrix to the Wigner distribution in phase space
\be
f({p,R},t)=\sum\limits_q {\rm e}^{i{q R}} f({p,q},t) ,
\label{1b}
\ee
where $R$ is the spatial coordinate.

The relaxation process in Eq. (\ref{1}) tends to 
establish a local equilibrium. Following Mermin, the local equilibrium can be characterized by a local variation of the chemical potential. Up to linear order it can be written as
\be
\left\langle p\!+\!{q\over 2}\right|\hat \rho^{\rm l.e.}
\left|p\!-\!{q\over 2}\right\rangle
=f_p \delta(q)
\!-\!{f_{p\!+\!\frac{q}{2}}\!-\!f_{p\!-\!\frac{q}{2}} \over \varepsilon_{p\!+\!\frac{q}{2}}\!-\!\varepsilon_{p\!-\!\frac{q}{2}}}\delta \mu(q,t)
\label{2}
\ee
with the Fermi-Dirac distribution $f_p=f_{\rm FD}\left ({\varepsilon_p}\right )$ and the kinetic energy
$\left\langle p+{q\over 2}\right|\hat{\cal E}\left|p-{q\over 2}\right\rangle=\varepsilon_p\delta(q)$.
The delta function $\delta(q)$ expresses that the corresponding terms are diagonal in momentum or translational invariant. The local deviation $\delta \mu$ from the global chemical 
potential is selected so that the density of particles is
conserved at each point and time instant.
Momentum and energy conservations can be included, too, leading to slightly 
more complicated formulas \cite{MF99,AA02}. 

Let us now specify $\delta\mu$ from the condition of density conservation
which requires that the local equilibrium distribution yields the
same density as the actual distribution, 
$n=\sum_p f=\sum_p f^{\rm l.e.}$. From Eq. (\ref{2}) the deviation from the 
density can be expressed as 
\be
\delta n(q,t)=\tilde \Pi^{\rm RPA}(t,\omega=0) \delta \mu(q,t),
\ee
where $\tilde \Pi^{\rm RPA}(t,\omega=0)$ is the  
polarization in random-phase approximation (RPA) 
at zero frequency with material 
parameters like electron density and the effective 
temperature fixed to the value at time $t$. Note that 
this RPA-polarization enters only the right hand side of 
(\ref{1}). The actual polarization which we will 
derive from (\ref{1}) is more complex.

The induced potential is given by the Poisson equation
\be
V^{\rm ind}({q},t)&=&\langle p+{q\over 2}|\hat {V}^{\rm ind} |p-{q\over 2}\rangle 
={e^2\over\epsilon_0 q^2} \delta n(q,t).
\ee

Now we can solve the kinetic equation (\ref{1}) for a small
external perturbation $V^{\rm ext}$. We split the distribution 
into the homogeneous part and a small inhomogeneous 
perturbation, $f(p,q,t)=f_p(t)+\delta f(p,q,t)$. Neglecting the 
time derivative of $f_p$ compared to $\delta f$ we can solve 
Eq. (\ref{1}) up to linear order in $V^{\rm ext}$
\be
&&\delta f(p,q,t)=i\!\int\limits_{t_0}^t \! dt' 
\exp{\left [\left (i \varepsilon_{p+\frac q 2}\!-\!i
\varepsilon_{p-\frac q 2}\!+\!\frac{1}{\tau}\right )(t'-t)\right ]}
\nonumber\\
&&\times
\Biggl \{
\left[f_{p+\frac q 2}(t')-f_{p-\frac q 2}(t')\right]
\left [{e^2\over\epsilon_0 q^2} \delta n(q,t')+
V_q ^{\rm ext}(t')\right ]
\nonumber\\
&&
+{1 \over i \tau \tilde \Pi^{\rm RPA}(t',0)}
{f_{p+\frac q 2}(t')-f_{p-\frac q 2}(t') \over 
\varepsilon_{p+\frac q 2}-\varepsilon_{p-\frac q 2}}
\delta n(q, t')
\Biggr \} 
\label{f}
\ee
where the perturbation starts at $\delta f(p,q,t_0)=0$.
 
The density response, $\chi(t,t')$, defined by 
\be
\delta n(q,t)=\int\limits_{t_0}^t dt' \chi(t,t')V_q ^{\rm ext}(t')
\label{dn}
\ee
follows from Eq. (\ref{f}) by integrating over $p$ with the result
\be
\chi(t,t')\!=\!\Pi(t,t')\!+\!\!
\int\limits_{t'}^t \! \! d {\bar t} 
\left [ \Pi(t,{\bar t}) {e^2\over\epsilon_0 q^2} \!+\! 
I(t,{\bar t})\right ] \chi({\bar t},t').
\label{chi}
\ee
Here Mermin's correction is represented by the term
\be
I(t,t')={i } \sum\limits_p {f_{p\!+\!\frac q 2}(t')\!-\!f_{p\!-\!\frac q 2}(t')\over \varepsilon_{p\!+\!\frac q 2}- \varepsilon_{p\!-\!\frac q 2}}{{\rm e}^{\left (i \varepsilon_{p\!+\!\frac q 2}-i \varepsilon_{p\!-\!\frac q 2}\!+\!\frac{1}{\tau}\right )(t'\!-\!t)}\over  \tau \tilde \Pi^{\rm RPA}(t',0)}.
\nonumber\\
&&
\label{i}
\ee
The polarization,
\be
\Pi(t,t')\!=\!i\! \sum\limits_p [f_{p\!+\!
\frac q 2}(t')\!-\!f_{p\!-\!\frac q 2}(t')]
{\rm e}^{\left (i \varepsilon_{\!\!p\!+\!\frac q 2}-i 
\varepsilon_{\!\!p\!-\!\frac q 2}\!+\!\frac{1}{\tau}\right )
(t'\!-\!t)},
\label{pi}
\ee
describes the response of the system with respect to the induced field in contrast to the response function (\ref{dn}) which describes the response of the system with respect to the external field.

To link the derived formulas with familiar results we note that for time-independent $f_p$, the Fourier transform of the polarization with respect to the difference time
\be
\tilde \Pi(T,\omega)=\int d \tau {\rm e}^{i \omega \tau}\Pi(T+\frac{\tau}{2},T-\frac{\tau}{2})
\label{pix}
\ee
becomes the standard RPA polarization with the 
frequency argument shifted by the inverse relaxation time
\be
\tilde \Pi(T,\omega)=\tilde \Pi^{\rm RPA}
\left(\omega+{i\over \tau}\right).
\ee
Moreover, in the limit $t+t'\to\infty$ the solution of Eq. (\ref{chi}) approaches the familiar form $\chi=\Pi^{\rm M}/(1-{e^2\over\epsilon_0 q^2} \Pi^{\rm M})$ where the Mermin-Das polarization reads \cite{Mer70}
\be
\Pi^{\rm M}={\tilde \Pi^{\rm RPA}(\omega+{i\over \tau})\over 
1-i(1+{1\over \omega \tau}) \left (1-{\tilde \Pi^{\rm RPA}
(\omega+{i\over \tau})\over\tilde \Pi^{\rm RPA}(0)}\right )}.
\ee
Eq. (\ref{chi}) describes how the Mermin
susceptibility is formed after a fast release of free carriers.

In the further analysis we will follow closely the experimental way of analyzing the two-time response function \cite{HKTBLVHKA05}. The pump pulse is creating charge carriers in the conduction band at time $t_0$ and the probe pulse is sent a time $t_D$ later. The time delay after this probe pulse  $T=t-t_D-t_0$ is then Fourier transformed into frequency. Of course everything starts at $t_0$ when the pump pulse has created the carriers in the conduction band, i.e. the maximum delay is $T \le t-t_0$. The frequency-dependent inverse dielectric 
function associated with the actual time $t$ thus 
reads
\be
z(t)={1 \over \epsilon(\omega,t)}-1=
{e^2\over\epsilon_0 q^2}\int\limits_{0}^{t-t_0} dT {\rm e}^{i \omega T}  \chi(t,t-T).
\label{ft}
\ee
This is exactly the one-sided 
Fourier transform introduced in Ref. \cite{SBHHH94}. 
It is worth noting that Wigner's form of the Fourier 
transform (\ref{pix}) would just result in a factor of two 
in the time evolution of the dielectric function. 

Now we turn our attention to short times when the response
is built up. In general the integral equation (\ref{chi}) together 
with Eqs. (\ref{i}) and (\ref{pi}) has to be solved. The useful 
limit of long wave lengths $q\to 0$ offers an appreciable 
simplification in that the leading terms of (\ref{pi}) and 
(\ref{i}) are $\Pi(t,t')\approx {q^2 n(t') \over m} (t'-t) 
{\rm e} ^{t'-t\over \tau}$ and  $I(t,t')\approx {1\over \tau} 
{\rm e}^{t'-t\over \tau}$. \cite{comm} 
It is practical to transform the integral equation (\ref{chi}) for the response function directly into one for the inverse dielectric function, Eq. (\ref{ft}), by differentiating twice with respect to time
\be
&&
z''(t)
+({1\over \tau}\!-\!2 i \omega) 
z'(t)\!+\!\left [\omega_p^2(t)\!-\!\omega^2\!-\!i{\omega \over \tau} \right ] z(t)\!=\!-\omega_p^2(t)
\nonumber\\
&&z(t_0)=0;\qquad \left.
z'(t)\right|_{t=t_0}=0.
\label{diff1}
\ee
Here $z'(t)=d z(t)/dt$ and we have denoted the time-dependent plasma frequency by 
$\omega_p^2(t)={e^2 n(t)\over \epsilon_0 \epsilon_\infty m}$.

In our simplified treatment the density of electrons $n(t)$ is 
assumed to be a known function of the time. Due to Rabi
oscillations it is a nontrivial function of the laser pulse. If the
pulse is short compared to the formation of the collective 
mode, all details of the density build-up become unimportant
and we can approximate the time dependence of the density 
by a step function \cite{SBHHH94}. Then the plasma frequency 
is $\omega_p(t)=\omega_p \Theta(t-t_0)$, where $t_0$ 
is the time when the pulse reaches its maximum. For this 
approximation the differential equation (\ref{diff1}) 
can be solved analytically, yielding
\be
{1 \over\epsilon(\omega,t)}=1-
{\omega_p^2\over\Gamma} 
\int\limits_0^{t-t_0} d t' \,{\rm e}^{(i\omega-{1\over 2 \tau}) t'} \,\sin{(\Gamma t')},
\label{result}
\ee
where $\Gamma=\sqrt{\omega_p ^2-{1\over 4 \tau ^2}}$. 
The integral (\ref{result}) can be expressed in terms of elementary 
functions, but we find the integral form more 
transparent. 

Formula (\ref{result}) has been derived from Mermin's 
density-conserving approximation. The long-time limit 
yields the Drude formula
\be
\lim\limits_{t\to\infty} {1 \over \epsilon}=
1-{\omega_p ^2\over \omega_p ^2-
\omega(\omega+{i\over \tau})}.
\ee
This obvious limit is not so 
easy to achieve within short-time expansions. For example 
the approximate result from quantum kinetic theory presented 
in \cite{SBHHH94} gives the long-time limit of 
the form $1-\omega_p^2/[\omega_p^2-(\omega+i/\tau)^2]$. 

\begin{figure}[h]
\psfig{file=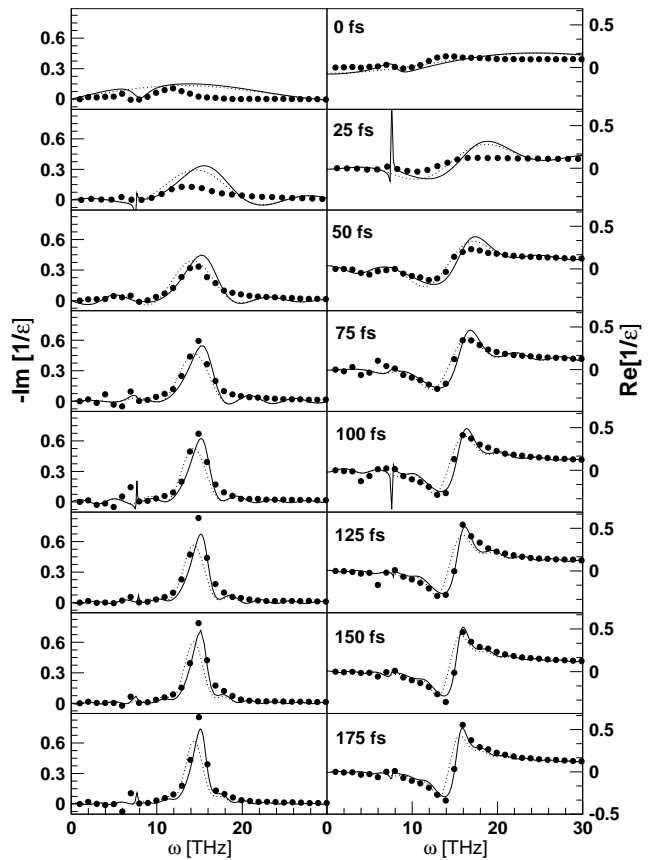,width=8.5cm}
\caption{The time evolution  
of the inverse dielectric function in 
GaAs. The labels from top to bottom denote the time $t$. The pumb pulse was at $t_0=-40$fs and the probe pulse has a full-width at half-maximum of 27$fs$. Circles are data from \protect\cite{HTBBAL01,HTBL02}, 
solid lines show the electronic part with intrinsic polarization 
(\protect\ref{form}) and dotted lines the electronic part 
alone (\protect\ref{result}). The plasma frequency is given by $\omega_p=$14.4 THz and the relaxation time is $\tau=85$ fs.  
}
\label{GaAs_1}
\end{figure}

In spite of the numerous approximations, Eq. (\ref{result}) 
fits well the experimental data, as shown in Fig. ~\ref{GaAs_1}
for the polar semiconductor GaAs.

The resonance visible at about 8~THz is due to optical 
phonons. This feature is easily accounted for, see Fig.~\ref{GaAs_1}, 
by adding the 
intrinsic contribution of the crystal lattice \cite{HTBL02}
\be
\epsilon_{\rm GaAs}(\omega,t)=\epsilon_\infty \left (\epsilon(\omega,t)+
{\omega_{\rm LO}^2-\omega_{\rm TO}^2 \over \omega_{\rm TO}^2-\omega^2-i \gamma \omega} \right ),
\label{form}
\ee
with the longitudinal and transversal optical frequencies 
$\omega_{\rm LO}/2\pi=8.8$ THz and $\omega_{\rm TO}/2\pi=
8.1$ THz, the lattice damping $\gamma=$0.2 ps$^{-1}$ and the
non-resonant background polarizability of the ion lattice $\epsilon_\infty=11.0$, see Ref.~\cite{HTBL02}. 

\begin{figure}[]
\psfig{file=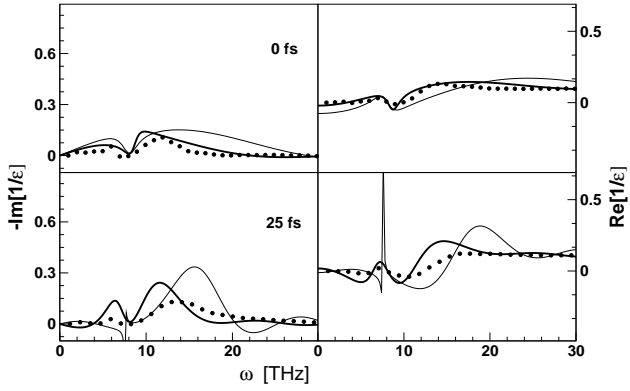,width=8cm}
\caption{The first two time evolution steps of Fig.~\protect\ref{GaAs_1} comparing the instant population of the conduction band (thin line) with a smooth transition (solid line).}
\label{GaAsimp}
\end{figure}

The formula (\ref{result}) results in a too fast build-up of the 
collective mode. This is seen in the second row of 
Fig.~\ref{GaAs_1} at the time $t=25$ fs. This is, however, just the time duration of the experimental pulse 
and consequently the time of populating the conduction band. 
Since we have approximated this by the instant 
jump, this discrepancy at short times around $25$~fs can be
expected. A more realistic smooth populating can be modeled by an 
arctan-function 
and then the numerical 
solution of (\ref{diff1}) improves the description as we see 
in Fig. \ref{GaAsimp}. Only the early development is plotted since the 
later stages agree with the simple estimate (\ref{result}). 

It is obvious that only the gross feature of formation of collective modes can be described by the mean-field fluctuations. For correlations beyond the mean field we should observe deviations. This is the case in the recently observed response of InP \cite{HKTBLVHKA05}, where a coupling between LO phonon and photon modes is reported. In Fig. \ref{InP} we see that the 
formation of the collective mode is delayed in the experiment 
during the first 100 fs when compared to the mean-field formula 
({\ref {result}). Such behavior is due to scattering processes neglected here.
At times 100~fs-200~fs the formation of collective modes is described again sufficiently well. 
For InP we use parameters according to \cite{HKTBLVHKA05}
namely $\omega_{\rm LO}/2\pi=10.3$~THz, 
$\omega_{\rm TO}/2\pi=9$~THz,  
$\gamma=0.2$~ps$^{-1}$, and $\epsilon_\infty=9.6$. 

\begin{figure}[]
\psfig{file=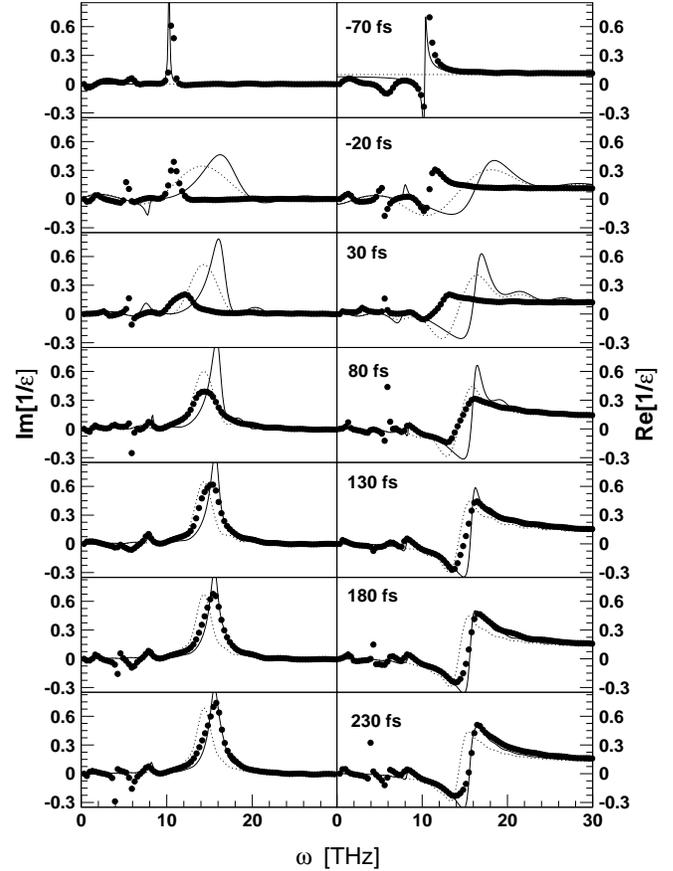,width=8.5cm}
\caption{Same as Fig.~\protect\ref{GaAs_1} but for InP. Circles are data from \protect\cite{HKTBLVHKA05}. 
The plasma frequency and 
the relaxation time are $\omega_p=$14.4 THz and $\tau=77$~fs, 
respectively and $t_0=-70$ fs.  
}\label{InP}\end{figure}

The aim of the present paper was to separate the gross feature of the 
formation of collective modes at transient times which are due to simple 
mean-field fluctuations. This has resulted in a simple analytic formula 
for the time dependence of the dielectric function. Subtracting this 
gross feature from the data allows one to extract the effects which are due 
to higher-order correlations and which have to be simulated by quantum 
kinetic theory \cite{BVMH98,GBH99,VH00,KK04} and response functions 
with approximations beyond the mean field \cite{KB00}. These treatments are
numerically demanding such that analytic 
expressions for the time dependence of some variables \cite{MSL97a}
are useful for controlling the numerics.

To conclude, we have derived a simple analytic formula for
the formation of a collective mode. Being able to describe universal 
features of the formation 
of quasiparticles, the simplicity of the presented
result is extremely practical and offers a wide range of
applications. We see two prominent examples: (1) It could spare a lot of 
computational power to simulate ultrashort-time behavior of new
nano-devices and (2) it can help to understand and describe 
the formation of collective modes during nuclear collisions which are
not experimentally accessible in the early phase of collision. 

We thank A. Huber and R. Leitenstorfer for providing us with the 
experimental data and H. N. Kwong for enlightening and clarifying 
discussions.


\begin{thebibliography}{21}
\bibitem[{com()}]{comm}
\bibinfo{note}{The long wave-length limit corresponds here to the
  quasiclassical limit, of course. Though we could have started directly with
  the quasiclassical kinetic equation, we prefer to keep here the more general
  quantum formalism since the same procedure can be applied for finite systems
  leading then to explicit quantum corrections.}
\end{thebibliography}

\end{document}